\pdfoutput=1
\documentclass[aip,apl,amsmath,amssymb,reprint]{revtex4-2}

\usepackage{graphicx}
\usepackage{dcolumn}
\usepackage{bm}
\usepackage{xcolor}
\usepackage[hidelinks]{hyperref}
\newcommand{\SMlink}[1]{\hyperlink{SM}{\textcolor{blue}{#1}}}
\makeatletter
\def\active@comma{,}
\makeatother

\graphicspath{{figures/}}

\newcommand{\dB}{\Delta B}

\newcommand{\Meff}{M_{\mathrm{eff}}}
\newcommand{\Bres}{B_{\mathrm{res}}}
\newcommand{\BT}{B_{\mathrm{T}}}
\newcommand{\TN}{T_{\mathrm{N}}}
\newcommand{\alphaeff}{\alpha_{\mathrm{eff}}}

\begin{document}

\title[Low magnon damping of a EuSe film down to millikelvin temperatures]{Low magnon damping of a EuSe film down to millikelvin temperatures}

\author{Rostyslav~O.~Serha}
\email{rostyslav.serha@univie.ac.at}
\affiliation{Faculty of Physics, University of Vienna, Boltzmanngasse 5, 1090 Vienna,~Austria}

\author{Jan~Hofer}
\affiliation{Faculty of Physics, University of Vienna, Boltzmanngasse 5, 1090 Vienna,~Austria}

\author{Khrystyna~O.~Levchenko}
\affiliation{Faculty of Physics, University of Vienna, Boltzmanngasse 5, 1090 Vienna,~Austria}

\author{Andreas~Ney}
\affiliation{Institute of Semiconductor and Solid State Physics, Johannes Kepler University Linz, \mbox{Altenberger Stra{\ss}e 69, 4040 Linz, Austria}}

\author{Sergej~O.~Demokritov}
\affiliation{Institute of Applied Physics, University of M{\"u}nster, Corrensstra{\ss}e 2-4, 48149 M{\"u}nster, Germany}

\author{Gunther~Springholz}
\affiliation{Institute of Semiconductor and Solid State Physics, Johannes Kepler University Linz, \mbox{Altenberger Stra{\ss}e 69, 4040 Linz, Austria}}

\author{Andrii~V.~Chumak}
\email{andrii.chumak@univie.ac.at}
\affiliation{Faculty of Physics, University of Vienna, Boltzmanngasse 5, 1090 Vienna,~Austria}

\date{\today}

\begin{abstract}
The rapid development of quantum magnonics over the past decade has created a demand for magnetic materials that keep their magnon losses low at millikelvin temperatures and add functionality beyond that of the established iron garnets and ferromagnetic metals. The europium chalcogenides are natural candidates --- they are magnetic semiconductors whose moment of $7\,\mu_{\mathrm{B}}$ per Eu$^{2+}$ ion stems from a half-filled $4f$ shell with zero orbital momentum, so that spin-orbit coupling and intrinsic damping are expected to be weak. Here we characterize an epitaxial film of the metamagnet EuSe on diamagnetic BaF$_2$ by magnetometry and broadband ferromagnetic resonance from 10\,K down to 30\,mK. EuSe combines a saturation magnetization of 1100\,kA/m with antiferromagnetic, ferrimagnetic and ferromagnetic phases that are reached with fields below 0.25\,T. In contrast to a Gilbert-damped ferromagnet, the linewidth of the field-aligned film decreases with increasing frequency and field, and below 750\,mK it no longer depends on temperature. At 30\,mK, 68\,GHz and 1.8\,T, it reaches 5.5\,mT, close to the values of bulk single crystals, which corresponds to a magnon lifetime of 2\,ns and to an upper bound of $1.1\times10^{-3}$ for the Gilbert damping parameter. These properties make EuSe and the europium chalcogenides promising materials for hybrid quantum systems and quantum magnonics.
\end{abstract}

\maketitle

Quantum magnonics aims to use magnons as carriers of quantum information and as a coherent interface between superconducting qubits and microwave photons.\cite{LachanceQuirion2019,Yuan2022,Chumak2022} Its experiments are performed in dilution refrigerators, where the thermal magnon and phonon populations are frozen out, and they rely on established magnetic materials. Efficient quantum magnonic experiments require long magnon lifetimes and thus materials whose dominant dissipation channels are suppressed in the quantum limit.\cite{Serha2026APLM,Serha2026SciAdv,Serha2026} Insulating magnets are preferred, since they are free of conduction-electron losses, but their material base in the field is narrow, and only a few have been studied down to millikelvin temperatures.\cite{Serha2026APLM} Nearly all experiments use yttrium iron garnet (YIG),\cite{Huebl2013,Tabuchi2014,Tabuchi2015,LachanceQuirion2020} which has the lowest known magnetic damping.\cite{Serha2026SciAdv} YIG films, however, are grown on gadolinium gallium garnet (GGG), a paramagnet whose large susceptibility below a few kelvin produces inhomogeneous stray fields and additional relaxation channels, so that the magnon losses of YIG films increase by up to an order of magnitude below 1\,K.\cite{Jermain2017,Mihalceanu2018,Kosen2019,Guo2022,Knauer2023,Serha2024,Serha2025,Schmoll2025,Schmoll2025LTP} A film grown on a diamagnetic substrate eliminates this stray-field mechanism, which for YIG required the design of a novel and complex diamagnetic substrate.\cite{Guguschev2025,Serha2026} Beyond the substrate, the field would benefit from insulating magnets with properties that YIG does not offer.

\begin{figure*}[t]
\centering
\includegraphics[width=\textwidth]{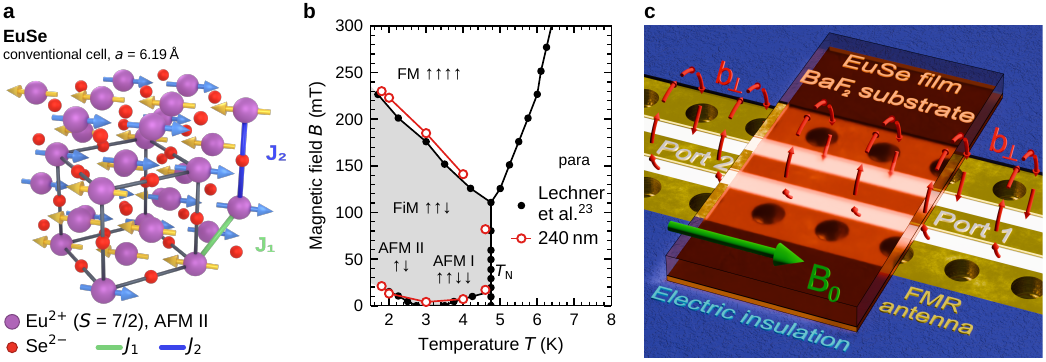}
\caption{\label{fig:1}(a) Rocksalt structure of EuSe with the conventional cubic cell outlined (space group $Fm\bar{3}m$, $a = 6.19$\,\AA{}). Arrows on the Eu$^{2+}$ ions ($S = 7/2$) show the $\uparrow\downarrow$ order of the ferromagnetic (111) sheets (AFM II), the ground state at zero temperature and zero field.\cite{Griessen1971,Rumpf2005,Lechner2005} The spins lie within the (111) planes,\cite{Wang2022} and the arrows point along $[1\bar{1}0]$, one of the directions in these planes. The ferromagnetic nearest-neighbor exchange $J_1$ (Eu-Eu distance $a/\sqrt{2}$, $J_1/k_{\mathrm{B}} = +0.22$\,K) and the antiferromagnetic next-nearest-neighbor exchange $J_2$ (distance $a$, through Se, $J_2/k_{\mathrm{B}} = -0.23$\,K)\cite{Sollinger2010} are marked once each. Reference~\onlinecite{Wang2022} uses the opposite labels for AFM I and AFM II. (b) Field and temperature phase diagram of EuSe. Black dots and lines are all data points and phase boundaries of a nearly unstrained, \textmu m thick EuSe film, redrawn from the $H$-$T$ phase diagram of Ref.~\onlinecite{Lechner2005} (obtained there from susceptibility measurements), with the antiferromagnetic (AFM I, AFM II), ferrimagnetic (FiM), field-aligned ferromagnetic (FM) and paramagnetic regions as labeled there. Red open circles mark the FiM to FM transition field $\BT$ and the AFM to FiM transition field of the 240\,nm film, likewise obtained from the maxima of the susceptibility of the VSM data. (c) FMR setup geometry. The EuSe film on its BaF$_2$ substrate lies face down on a coplanar waveguide, the static field $B_0$ is applied in the film plane along the signal line, the microwave field $b_\perp$ of the waveguide excites the precession, and the complex transmission $S_{21}$ between the two ports is recorded at fixed frequency while $B_0$ is swept.}
\end{figure*}

The europium monochalcogenides EuO, EuS, EuSe and EuTe are attractive in this respect.\cite{Wachter1979,Mauger1986} They are magnetic semiconductors in which each Eu$^{2+}$ ion carries a moment of $7\,\mu_\mathrm{B}$. The band gap of EuSe is 2.05\,eV at 1.7\,K.\cite{Kirchschlager2004} The half-filled $4f^{7}$ shell has $S = 7/2$ and $L = 0$, so that spin-orbit coupling, and with it the intrinsic damping, is expected to be weak. The localized $4f$ electrons make them the reference Heisenberg magnets on a rocksalt lattice. The family spans the ferromagnets EuO and EuS, the metamagnet EuSe and the antiferromagnet EuTe. It has long served spintronics as spin-filter tunnel barriers,\cite{Moodera1988,Moodera1993,Santos2004} shows large magnetoresistance at its magnetic transitions\cite{Shapira1974} and is exchange coupled to topological insulators at interfaces.\cite{Katmis2016,Wang2023} In EuSe, the ferromagnetic nearest-neighbor exchange $J_1$ and the antiferromagnetic next-nearest-neighbor exchange $J_2$ nearly cancel,\cite{Sollinger2010} so that several stacking orders of the ferromagnetic (111) sheets are almost degenerate (see Fig.\,\ref{fig:1}\,(a)). Its saturation magnetization of up to 1100\,kA/m is more than five times that of YIG at low temperatures.\cite{Serha2025PRB} Below the N\'eel temperature $\TN \approx 4.6$\,K, the spins order antiferromagnetically (AFM I, AFM II), a small in-plane field stabilizes the ferrimagnetic $\uparrow\uparrow\downarrow$ phase, and a field of about 0.2\,T aligns all spins (see Fig.\,\ref{fig:1}\,(b)).\cite{Schwob1966,Fischer1969,Griessen1971,Callen1977,Lechner2005,Diaz2010,Rumpf2005} In zero field below $\TN$, EuSe thus carries no net moment and no stray field, which is attractive next to superconducting elements, whereas in the field-aligned state magnetostatic surface spin waves would propagate about 35 times faster than in YIG films of the same thickness at the same frequency.\cite{Serha2025PRB,DamonEshbach1961} The same near-cancellation makes the magnetic order tunable by strain,\cite{Lechner2005} hydrostatic pressure\cite{Goncharenko1998,Sollinger2010} and light.\cite{Henriques2018,vanKooten2022} Nearly unstrained EuSe films grow by molecular beam epitaxy on BaF$_2$(111), a diamagnetic insulator with almost the same lattice constant.\cite{Kirchschlager2004,Lechner2005,Wang2022}

The magnon damping of EuSe, a key parameter for its application in quantum magnonics, remains largely unexplored. Ferromagnetic resonance (FMR), the standard probe of this damping, has so far been applied to EuSe only on bulk single-crystal spheres and only above 1\,K, where linewidths of 2 to 7.5\,mT were found at 21 to 24\,GHz,\cite{Brown1968,Everett1971} together with $g = 1.98 \pm 0.01$.\cite{Brown1968} Comparable early data exist for EuS and EuO.\cite{vonMolnar1965,Dillon1964} Neither films nor the temperature range below 1\,K, which is relevant for quantum magnonics, have been studied.

\begin{figure*}[t]
\centering
\includegraphics[width=\textwidth]{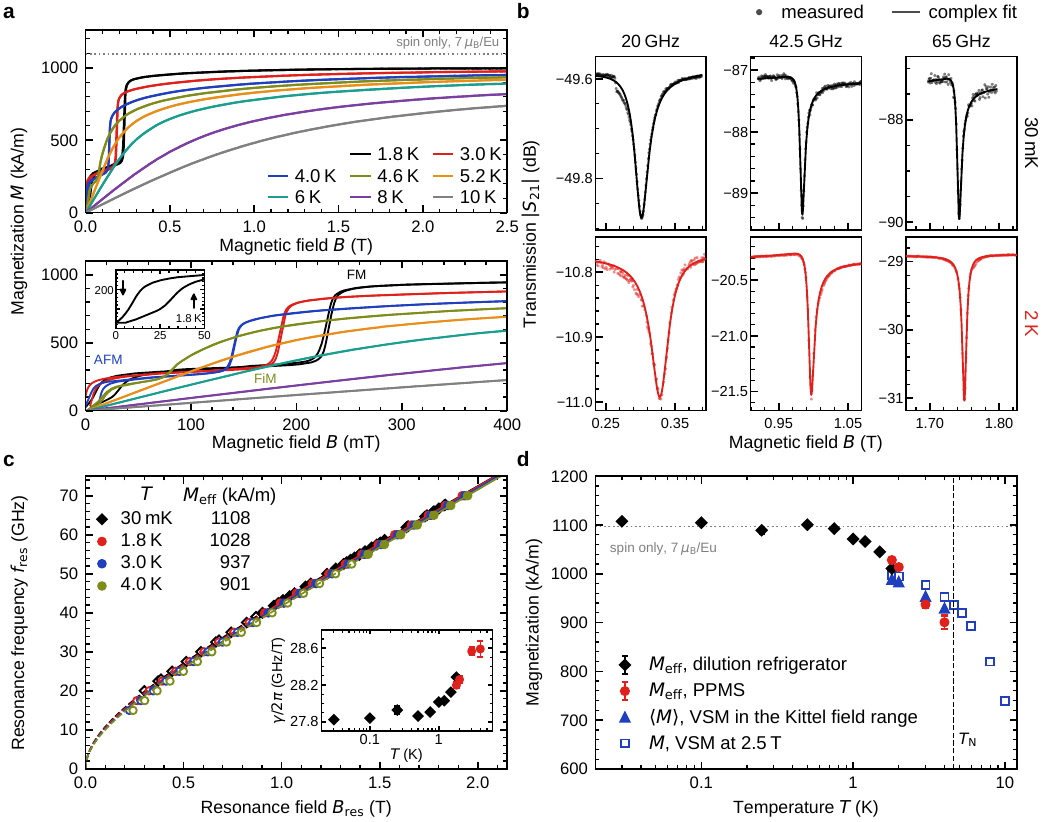}
\caption{\label{fig:2}(a) Magnetization of the 240\,nm film versus in-plane field at 1.8\,K to 10\,K (VSM, up and down sweeps to 2.5\,T). The lower part shows the same data below 400\,mT with the antiferromagnetic (AFM), ferrimagnetic (FiM) and field-aligned ferromagnetic (FM) regions. Its inset shows the hysteretic transition between the AFM and FiM states at 1.8\,K, with arrows for the up and down sweep. (b) Measured FMR transmission $|S_{21}|$ at 20\,GHz, 42.5\,GHz and 65\,GHz, 30\,mK in the upper row and 2\,K in the lower row. Symbols are the measured traces, lines the complex Lorentzian fits. Every sub-panel spans 160\,mT, and both rows share the field axis of each frequency. The different transmission levels of the two rows reflect the different microwave lines of the dilution refrigerator and the PPMS. (c) Resonance frequency versus resonance field at 30\,mK, 1.8\,K, 3.0\,K and 4.0\,K with in-plane Kittel fits, Eq.~\eqref{eq:kittel}, which give the effective magnetization $\Meff$ listed in the panel. Filled symbols are included in the fits. Open symbols lie below the cutoff frequency of 40\,GHz (55\,GHz at 4.0\,K). Dashed lines extrapolate the fits below the lowest fitted frequency. The inset shows the gyromagnetic ratio $\gamma/2\pi$ of all Kittel fits versus temperature, with the symbols of (d). (d) Effective magnetization $\Meff$ from the Kittel fits up to 4\,K, diamonds from the dilution refrigerator, circles from the PPMS, and VSM magnetization from (a), averaged over the field range of each PPMS Kittel fit (triangles) and at 2.5\,T (open squares), versus temperature. The dotted line is the spin-only saturation magnetization for $7\,\mu_{\mathrm{B}}$ per Eu, the dashed line marks $\TN$. Where no error bars are visible, they are smaller than the symbols.}
\end{figure*}

Here we present broadband FMR of an epitaxial EuSe film on BaF$_2$ from 15 to 70\,GHz and from 10\,K down to 30\,mK, together with magnetometry of the metamagnetic transitions. The 240\,nm thick EuSe film was grown by molecular beam epitaxy on a 600\,µm thick BaF$_2$(111) substrate, as described in Ref.~\onlinecite{Lechner2005}.
The static magnetization was measured by vibrating sample magnetometry (VSM) in a physical property measurement system (PPMS, Quantum Design) between 1.8 and 10\,K with the field in the film plane, in up and down sweeps up to 2.5\,T. The diamagnetic background of the substrate was subtracted using the signal measured at 300\,K, and the magnetization was normalized to the nominal film volume. The transition fields were taken at the maxima of the susceptibility $dM/dH$.

For FMR spectroscopy, a 3\,mm wide piece of the film was placed face down on a coplanar waveguide, covering its 200\,\textmu m wide signal line, and the complex transmission $S_{21}$ was recorded with a vector network analyzer (VNA, Anritsu MS4647B) at fixed frequencies while the field, applied in the film plane along the signal line, was swept (see Fig.\,\ref{fig:1}\,(c)). Between 1.8 and 10\,K, the measurements were performed in the PPMS at frequencies from 15 to 70\,GHz, and from 30\,mK to 1.8\,K in a dilution refrigerator (Bluefors LD) with a superconducting vector magnet (American Magnetics 430) at frequencies up to 67.7\,GHz. The microwave power was set to $-10$\,dBm at the VNA output and about $-70$\,dBm at the sample in the dilution refrigerator, while the field was ramped slowly to avoid heating of the sample. In the PPMS the power was at least one order of magnitude below the onset of nonlinear broadening determined in a power sweep at 1.8\,K.

Each FMR trace was fitted with a complex Lorentzian on a complex linear background, which yields the resonance field $\Bres$ and the loaded full linewidth at half maximum (FWHM) $\dB_{\mathrm{L}}$. Up and down field sweeps were averaged where available. Since the magnon mode acts as a resonator side-coupled to the waveguide, the loaded linewidth contains the radiative contribution of this coupling.\cite{Schoen2015} Following the analysis of notch-type resonators,\cite{Probst2015} the intrinsic linewidth is obtained from the depth $D$ of the symmetrized transmission dip, in dB, as $\dB = \dB_{\mathrm{L}}\,10^{-D/20}$. The correction ranges from a few percent for the weakly coupled traces at low fields to about one third for the most strongly coupled traces at high frequencies. All linewidths discussed below are intrinsic FWHM values in field.

\begin{figure*}[t]
\centering
\includegraphics[width=\textwidth]{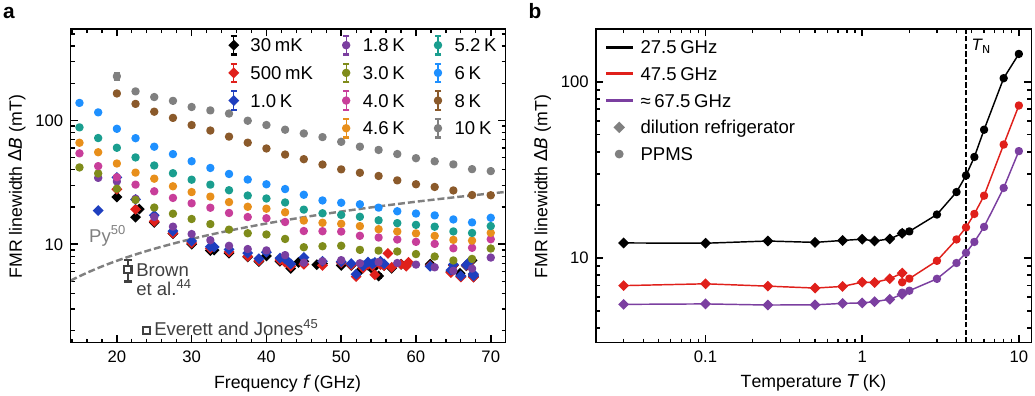}
\caption{\label{fig:3}(a) Intrinsic FMR linewidth versus frequency at 30\,mK, 500\,mK and 1.0\,K (dilution refrigerator, diamonds) and at 1.8, 3.0, 4.0, 4.6, 5.2, 6, 8 and 10\,K (PPMS, circles). Because the frequencies were chosen at transmission maxima of the microwave lines, they differ by up to 0.2\,GHz between temperatures. Open squares are the linewidths of bulk EuSe spheres as reported, 5 to 7.5\,mT at 21.48\,GHz and 1.3\,K\cite{Brown1968} and 2\,mT at 24\,GHz and 1.15\,K,\cite{Everett1971} measured at internal fields of 0.3 to 0.4\,T. Both were recorded as field derivatives of the absorption. If these widths are peak-to-peak values, the corresponding Lorentzian FWHM is larger by a factor of $\sqrt{3}$. The dashed line is the calculated Gilbert linewidth $\dB = 4\pi\alpha f/\gamma$ of Permalloy with the bulk damping $\alpha = 5.4\times10^{-3}$ at 10\,K of Ref.~\onlinecite{Ney2026} and the typical gyromagnetic ratio $\gamma/2\pi = 29.4$\,GHz/T,\cite{Shaw2013} without inhomogeneous broadening. (b) Linewidth versus temperature at 27.5, 47.5 and 67.5\,GHz from 30\,mK to 10\,K, diamonds from the dilution refrigerator, circles from the PPMS. At 1.8\,K, the linewidth was measured in both setups. Error bars are the statistical uncertainties of the fits. Where no error bars are visible, they are smaller than the symbols.}

\end{figure*}

Figure\,\ref{fig:2}\,(a) shows the static magnetization of the film versus in-plane field. Up to 4.6\,K, it exhibits the two metamagnetic steps of EuSe. At a few tens of millitesla, the antiferromagnetic order changes into the ferrimagnetic $\uparrow\uparrow\downarrow$ order, in which the magnetization stays on a plateau of about one third of its field-aligned value and no FMR signal is detected between 15 and 70\,GHz, as expected. At the transition field $\BT$, the magnetization jumps into the field-aligned ferromagnetic state. With increasing temperature, $\BT$ shifts from 230\,mT at 1.8\,K to 141\,mT at 4\,K, and both transition fields, $B_\mathrm{T}$ and the field of the antiferromagnetic to ferrimagnetic transition, shown as red open circles in Fig.\,\ref{fig:1}\,(b), follow the phase diagram of a nearly unstrained, µm-thick EuSe film\cite{Lechner2005} within 10\,mT. Because the phase boundaries are strain dependent,\cite{Lechner2005} this means that no significant strain is present in the film. A hysteresis between up and down sweeps appears only at the antiferromagnetic to ferrimagnetic transition, and at zero field the magnetization returns to about 3\,\% of its value at 2.5\,T, so that the film nearly recovers the zero-moment antiferromagnetic state. Above $\TN$, the steps vanish, and the magnetization grows continuously with the field. At 1.8\,K and 2.5\,T, it reaches 997\,kA/m, which is 91\,\% of the spin-only saturation magnetization. All FMR data below $\TN$ refer to the field-aligned ferromagnetic state.

Figure\,\ref{fig:2}\,(b) shows typical transmission traces at 30\,mK and 2\,K. At all frequencies, the resonance is a single dip, which the complex Lorentzian fit reproduces.
From 30\,mK to 2\,K, the resonance shifts to higher fields as the magnetization decreases, and the line broadens at low but not at high frequencies.

The resonance fields follow the Kittel dispersion of an in-plane magnetized film,\cite{Kittel1948}
\begin{equation}
f_{\mathrm{res}} = \frac{\gamma}{2\pi}\sqrt{\Bres\left(\Bres + \mu_0 \Meff\right)},
\label{eq:kittel}
\end{equation}
where $\gamma = g\mu_{\mathrm{B}}/\hbar$ is the gyromagnetic ratio with the Land\'e factor $g$, and the effective magnetization $\Meff$ contains the saturation magnetization and the anisotropy fields (see Fig.\,\ref{fig:2}\,(c)). Because the magnetization still grows with the field close to $\BT$ (see Fig.\,\ref{fig:2}\,(a)), only resonances above a cutoff frequency of 40\,GHz (55\,GHz at 4\,K) are fitted. At millikelvin temperatures, the fit yields $\gamma/2\pi = 27.82$\,GHz/T, which corresponds to $g = 1.99$ and agrees with the bulk value $g = 1.98 \pm 0.01$,\cite{Brown1968} and $\Meff = 1108$\,kA/m, within 1\,\% of the spin-only saturation magnetization of 1097\,kA/m obtained for $7\,\mu_{\mathrm{B}}$ per Eu ion and the lattice constant $a = 6.186$\,\AA{}.\cite{Wang2022}
The film thus carries the full Eu$^{2+}$ moment, and the net anisotropy field is small within the accuracy of the in-plane fit. Up to 500\,mK, $\Meff$ is constant within 2\,\% (Fig.\,\ref{fig:2}\,(d)), and the values of the two setups agree within 2\,\% at 1.8\,K. At higher temperatures, $\Meff$ decreases together with the VSM magnetization averaged over the fit range, but about twice as fast. This deviation, and the apparent increase of the fitted $\gamma/2\pi$ to 28.6\,GHz/T between 1.8 and 4\,K (inset of Fig.\,\ref{fig:2}\,(c)), result from the residual field dependence of the magnetization in the field-aligned state, which the Kittel dispersion of a saturated ferromagnet does not include. Above 4\,K, the magnetization depends strongly on the field, and a superparamagnetic phase of partly ordered spins has been reported,\cite{Lechner2005} so that no Kittel analysis is performed.

Figure\,\ref{fig:3}\,(a) presents the central result of this work, the intrinsic FMR linewidth $\dB$ versus frequency from 30\,mK to 10\,K. For comparison, the dashed line shows the Gilbert linewidth of Permalloy at 10\,K, which increases linearly with frequency.\cite{Ney2026} In contrast, the linewidth of EuSe below $\TN$ decreases with increasing frequency and approaches a constant value above about 45\,GHz. At 30\,mK, it falls from 24\,mT at 20\,GHz, where the resonance field lies close to the FiM to FM transition, to 5.45\,mT at 67.7\,GHz. At high frequencies, the linewidths below 750\,mK coincide with the 30\,mK data, and the 1.8\,K data lie only slightly higher. Above $\TN$, the linewidth increases by an order of magnitude at low frequencies but still decreases with frequency, even at high resonance fields, which is consistent with a further alignment of the spins by the field, since above $\TN$ the magnetization does not saturate (see Fig.\,\ref{fig:2}\,(a)). A 500\,nm thick EuSe film shows a similar decrease of the linewidth with increasing frequency from 4 to 16\,K (see \SMlink{supplementary material}).

Gilbert damping cannot describe this behavior. Below 2\,K, the additional broadening at low frequencies decreases with the distance of the resonance field from $\BT$ and vanishes within the scatter of about 1\,mT between neighboring frequencies for resonance fields above about 1.2\,T. This points to a contribution of the nearby metamagnetic transition that lies beyond the macrospin Kittel picture. At high fields, the remaining linewidth is frequency independent within this scatter, so that no Gilbert slope can be resolved. If the entire linewidth of 5.45\,mT at 67.7\,GHz were of Gilbert-damping origin, the damping parameter would be $\alphaeff = \gamma\dB/(4\pi f) = 1.1\times10^{-3}$, which is therefore an upper bound for the Gilbert damping of the film. The magnon lifetime $\tau = 1/(\pi\Delta f)$ in the convention of Ref.~\onlinecite{Serha2026}, with the frequency linewidth $\Delta f = \dB\,\partial f_{\mathrm{res}}/\partial\Bres = 158$\,MHz, is 2\,ns.

At high frequencies, the linewidth of the film approaches the values of bulk spheres, 5 to 7.5\,mT at 21.48\,GHz and 1.3\,K\cite{Brown1968} and 2\,mT at 24\,GHz and 1.15\,K,\cite{Everett1971} which were, however, measured at internal fields of only 0.3 to 0.4\,T, where the film is several times broader. The strong broadening at low fields is thus specific to the film. In bulk EuSe, the linewidth depended strongly on the purity of the crystals. Spheres from undistilled material showed about 10\,mT, annealing narrowed the line by about a factor of two,\cite{Brown1968} and the narrowest lines were obtained after purification by fractional distillation.\cite{Everett1971} Impurities and lattice defects, together with the classical sources of inhomogeneous broadening in films such as strain and thickness variations, may therefore contribute to the linewidth of the film. Slowly relaxing impurities with relaxation time $\tau_{\mathrm{i}}$ add a linewidth proportional to $\omega\tau_{\mathrm{i}}/[1+(\omega\tau_{\mathrm{i}})^2]$, which decreases with increasing frequency once $\omega\tau_{\mathrm{i}} > 1$ and depends on temperature through $\tau_{\mathrm{i}}$.\cite{Clogston1955,Seiden1964} Such a contribution is consistent with the low-frequency linewidth, which increases between 30\,mK and 1\,K (see Fig.\,\ref{fig:3}\,(a)), whereas the high-frequency linewidth stays nearly constant. Compared with Permalloy at 10\,K, EuSe is broader below 30\,GHz but four times narrower at 66\,GHz, and near 70\,GHz its linewidth is still about 15 times larger than that of YIG films on diamagnetic garnet substrates at low temperatures.\cite{Serha2026}

Figure\,\ref{fig:3}\,(b) shows the linewidth versus temperature at three frequencies. Up to 750\,mK, where $\Meff$ begins to decrease, the linewidth at these frequencies stays within 4\,\% of its value at 30\,mK.
With increasing temperature, it broadens only moderately up to about 2\,K, and above 2\,K it rises steeply, together with the decrease of the magnetization in Fig.\,\ref{fig:2}\,(d). The strong broadening around $\TN$ resembles the broadening by critical spin fluctuations near a magnetic phase transition, observed in ultrathin Ni films at their Curie temperature.\cite{Li1990} Here, all resonance fields lie well above the metamagnetic transition, and the linewidth keeps increasing above $\TN$. At 5.2\,K and above, the magnetization shows no metamagnetic transition (see Fig.\,\ref{fig:2}\,(a)), so that the FMR probes paramagnons of the paramagnetic phase, polarized by the external field. The linewidths measured in both setups at 1.8\,K agree within 1\,mT at all three frequencies.

The temperature independence up to 750\,mK constrains the nature of the residual broadening. At 27.5\,GHz, the thermal occupation of the resonant magnon mode, $n = [\exp(hf/k_{\mathrm{B}}T) - 1]^{-1}$, rises from a negligible value at 30\,mK to 0.2 at 750\,mK, while the linewidth changes by less than 4\,\%, which argues against thermally populated magnons and temperature-dependent scattering processes as its dominant origin. Furthermore, the diamagnetic substrate rules out the mechanism that dominates the low-temperature damping of YIG films on paramagnetic GGG, where the inhomogeneous stray field of the partially magnetized substrate increases the FMR linewidth up to tenfold at millikelvin temperatures.\cite{Serha2025} Like YIG on a diamagnetic garnet substrate,\cite{Serha2026} EuSe on BaF$_2$ shows no damping increase down to 30\,mK. The residual linewidth at high fields, which depends neither on temperature nor on frequency, is consistent with an inhomogeneous distribution of local resonance fields.

In summary, the FMR linewidth of an epitaxial EuSe film on a diamagnetic BaF$_2$ substrate decreases upon cooling, is temperature independent at high frequencies up to 750\,mK and, in contrast to that of a Gilbert-damped ferromagnet, decreases with increasing frequency and field. At 68\,GHz and 30\,mK, it reaches 5.5\,mT, which corresponds to a magnon lifetime of 2\,ns and to an upper bound of $1.1\times10^{-3}$ for the Gilbert damping parameter. Above $\TN$, at 5.2\,K and higher temperatures, the resonance is that of paramagnons of the field-polarized paramagnetic phase, whose linewidth increases with temperature but still decreases with frequency. The effective magnetization equals the full spin-only moment. The remaining broadening is extrinsic, and the narrower lines of purified bulk crystals\cite{Everett1971} indicate that it can be reduced further by materials development, as the history of YIG shows.\cite{Serha2026SciAdv} In addition to low magnon losses, EuSe provides a zero-field state without net moment and stray field, while fields below 0.25\,T switch it between antiferromagnetic, ferrimagnetic and ferromagnetic phases. These properties make EuSe a candidate material for hybrid quantum systems and quantum magnonics, while the magnon dynamics of its antiferromagnetic and ferrimagnetic phases remains to be explored. More generally, the europium chalcogenides unite ferromagnetic, metamagnetic and antiferromagnetic magnon systems with semiconducting transport in one material class, which makes them a versatile platform for quantum magnonics and spintronics.

\section*{Supplementary Material}
See the \SMlink{supplementary material} for a comparison of the resonance fields and FMR linewidths of the 240\,nm film with those of a 500\,nm thick EuSe film.

\begin{acknowledgments}
This research was funded in whole or in part by the Austrian Science Fund (FWF) projects ParaMagnonics [10.55776/I6568] and TopMag [10.55776/ESP526]. SOD acknowledges support from the Deutsche Forschungsgemeinschaft (DFG, German Research Foundation), Project-ID 523303368. For open access purposes, the authors have applied a CC BY public copyright license to any author accepted manuscript version arising from this submission.
\end{acknowledgments}

\section*{Author Declarations}
\subsection*{Conflict of Interest}
The authors have no conflicts to disclose.
\subsection*{Author Contributions}
\textbf{Rostyslav O. Serha.} Conceptualization (lead), Investigation (equal), Formal Analysis (equal), Writing/Original Draft Preparation (lead), Writing/Review \& Editing (equal).
\textbf{Jan Hofer.} Investigation (equal), Formal Analysis (equal), Writing/Review \& Editing (equal).
\textbf{Khrystyna O. Levchenko.} Methodology (supporting), Writing/Original Draft Preparation (supporting), Writing/Review \& Editing (equal).
\textbf{Andreas Ney.} Resources (supporting), Writing/Review \& Editing (equal).
\textbf{Sergej O. Demokritov.} Resources (supporting), Formal Analysis (supporting), Writing/Review \& Editing (equal).
\textbf{Gunther Springholz.} Resources (lead), Investigation (supporting), Writing/Review \& Editing (equal).
\textbf{Andrii V. Chumak.} Supervision (lead), Funding Acquisition (lead), Writing/Review \& Editing (equal).

\section*{Data Availability Statement}
The data that support the findings of this study are available from the corresponding author upon reasonable request.

\bibliography{EuSe_APLQ_letter_AC_colleagues}

\clearpage
\onecolumngrid
\setcounter{figure}{0}
\renewcommand{\thefigure}{S\arabic{figure}}
\renewcommand{\theHfigure}{S\arabic{figure}}

\noindent\hypertarget{SM}{}{\raggedright\sffamily\bfseries\Large Supplementary Material\\ Low magnon damping of a EuSe film down to millikelvin temperatures\par}

\section*{Comparison of the 240\,nm and 500\,nm thick EuSe films}

\begin{figure}[h!]
\includegraphics{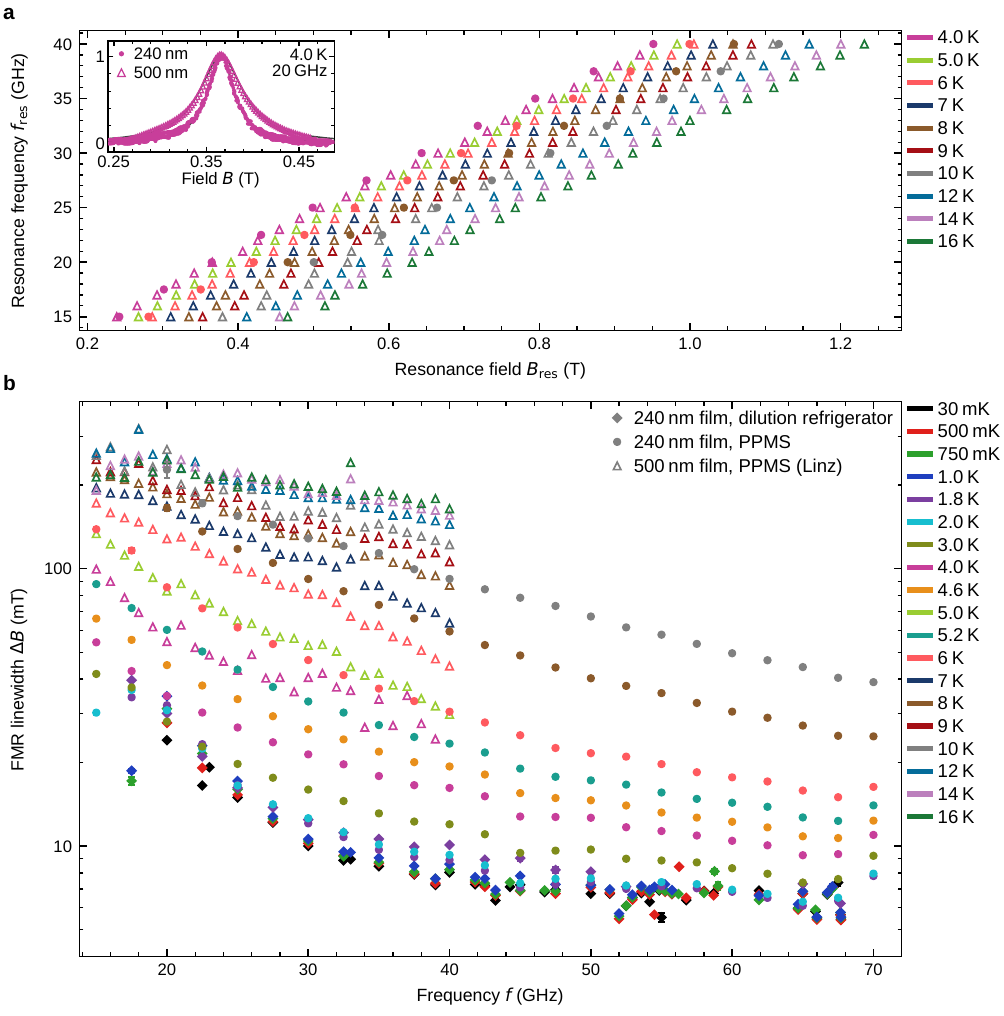}
\caption{\label{fig:S1}FMR of the two EuSe films on BaF$_2$(111), with the field applied in the film plane and the temperature given by the colors. (a) Resonance frequency versus resonance field of the 500\,nm film from 4 to 16\,K (open triangles, PPMS at Johannes Kepler University Linz) and of the 240\,nm film of the main text at the same temperatures, 4, 6, 8 and 10\,K (filled circles, PPMS), in the range of the 500\,nm data. The uncertainties of the 240\,nm resonance fields are smaller than the symbols. No Kittel fits are shown, because at these temperatures the magnetization still depends on the field (see main text). The inset compares the absorption lines of both films at 4\,K and 20\,GHz, normalized to the amplitude of a Lorentzian fit (lines). (b) FMR linewidth versus frequency. Filled symbols show the intrinsic FWHM linewidths of the 240\,nm film from the dilution refrigerator at 30\,mK, 500\,mK, 750\,mK, 1.0\,K and 1.8\,K (diamonds) and from the PPMS at 1.8 to 10\,K (circles), open triangles the FWHM linewidths of the 500\,nm film.}
\vspace{2pt}
\end{figure}

In thin films, the FMR linewidth contains inhomogeneous contributions from strain, thickness variations and defects, which depend on the growth and on the film thickness. A comparison of films with different thickness therefore allows contributions specific to a film to be distinguished from properties of the material. To test whether the frequency dependence of the linewidth reported in the main text is specific to the 240\,nm film and to the measurement setups, we compare it with broadband FMR data of a second EuSe film on BaF$_2$(111) with a thickness of 500\,nm.
This film was measured in a physical property measurement system (PPMS) at Johannes Kepler University Linz, with the field applied in the film plane, at frequencies from 15 to 40\,GHz in steps of 1\,GHz and at ten temperatures from 4 to 16\,K. For each frequency, the resonance field and the full width at half maximum (FWHM) of the resonance line were determined by fits of the FMR signal. The inset of Fig.\,\ref{fig:S1}\,(a) compares the absorptive part of its FMR spectrum at 4\,K and 20\,GHz with the absorption line of the 240\,nm film.
The 240\,nm film was measured from 15 to 70\,GHz, in the dilution refrigerator from 30\,mK to 1.8\,K and in the PPMS from 1.8 to 10\,K, as described in the main text. Its linewidths are the intrinsic FWHM values, corrected for the radiative contribution of the coupling to the waveguide.

Figure\,\ref{fig:S1}\,(a) shows the resonance frequency versus the resonance field of both films. With increasing temperature, the resonance of each film shifts to higher fields. At the temperatures that both films share, 4, 6, 8 and 10\,K, their resonance fields agree within 4\,\% at all common frequencies. The two films thus show the same dependence of the resonance frequency on the field within this accuracy, so that their linewidths are compared at nearly equal resonance fields. The difference in linewidth discussed below therefore does not stem from a different effective magnetization of the two films.

Figure\,\ref{fig:S1}\,(b) shows the linewidths of both films. As for the 240\,nm film, the linewidth of the 500\,nm film decreases with increasing frequency at every temperature. Between 20 and 40\,GHz, it falls by a factor of 2 to 3 in both films at all four common temperatures, so that the two films show a similar frequency dependence. Linear fits of the linewidth versus frequency, which would give the Gilbert damping parameter of a Gilbert-damped ferromagnet, yield negative slopes for the 500\,nm film at all temperatures.

With increasing temperature, the linewidth of the 500\,nm film increases, as does that of the 240\,nm film. At the highest frequencies it rises steadily from 4 to 16\,K, whereas at the lowest frequencies it no longer increases above about 10\,K. The decrease with frequency therefore becomes weaker at high temperatures, and at 16\,K the linewidth changes by only 23\,\% between 15 and 40\,GHz. It persists, however, up to 16\,K, far above the N\'eel temperature $T_\mathrm{N} \approx 4.6$\,K of EuSe, in agreement with the behavior of the 240\,nm film above $T_\mathrm{N}$ described in the main text.

At equal temperature, the linewidth of the 500\,nm film exceeds that of the 240\,nm film in the common frequency range by a factor of 1.1 to 1.9. This factor was evaluated at the frequencies measured in both films. It is largest at 4\,K, where it reaches 1.6 to 1.9, and decreases to 1.2 to 1.3 at 10\,K. The comparison is restricted to temperatures from 4 to 10\,K and to frequencies up to 40\,GHz. It covers neither the millikelvin range nor the frequencies above about 45\,GHz, where the linewidth of the 240\,nm film below $T_\mathrm{N}$ approaches a constant value (Fig.\,\ref{fig:3}\,(a) of the main text).

The decrease of the linewidth with increasing frequency is thus observed in two EuSe films of different thickness, measured in different setups and laboratories. This indicates that it is a property of EuSe films rather than of a particular sample or measurement setup.

\end{document}